# Radio Frequency Solid State Amplifiers

*J. Jacob*
ESRF, Grenoble, France

**Abstract**
Solid state amplifiers are being increasingly used instead of electronic vacuum tubes to feed accelerating cavities with radio frequency power in the 100 kW range. Power is obtained from the combination of hundreds of transistor amplifier modules. This paper summarizes a one hour lecture on solid state amplifiers for accelerator applications.

# 1 Introduction

The aim of this lecture was to introduce some important developments made in the generation of high radio frequency (RF) power by combining the power from hundreds of transistor amplifier modules. Such RF solid state amplifiers (SSA) were developed and implemented at a large scale at SOLEIL to feed the booster and storage ring cavities [1, 2]. At ELBE FEL four 1.3 GHz–10 kW klystrons have been replaced with four pairs of 10 kW SSAs from Bruker Corporation (now Sigmaphi Electronics, Haguenau/France), thereby doubling the available power [3]. The company Cryolectra GmbH (Wuppertal/Germany) delivers SSA solutions at various frequencies and power levels, such as a 72 MHz–150 kW SSA for a medical cyclotron [4].

Following a transfer of technology from SOLEIL, the company ELTA (Blagnac/France), a subsidiary of the French group AREVA, has delivered seven 352.2 MHz–150 kW RF SSAs to ESRF [5, 6]. These SSAs are used as an example to illustrate this technology. A compact SSA making use of a cavity combiner with fully planar RF amplifier modules that are suited for mass production is under development at ESRF: the aim is to reduce the required space and fabrication costs[1].

## 1.1 Typical radio frequency system layout

The typical layout of an RF transmitter powering an accelerating cavity in a particle accelerator is shown in Fig. 1. The low-level RF signal generated by the master source is distributed to all RF stations and other equipment such as, for example, the beam diagnostics systems. It is modulated in amplitude and phase by the low-level RF (LLRF) system, pre-amplified and amplified to high power by means of an RF power amplifier, which is the subject of this paper.

RF power amplifiers are often protected against reverse power by an isolator built with an RF power circulator that transmits the incident power to the accelerating cavity and deviates the reverse power into a high power load. The RF transmitter generally includes various auxiliaries, a power supply, a modulator, and fast interlock protection systems.

---

[1] This development has received funding from the EU as work package WP7 in the frame of the FP7/ESFRI/CRIPS project

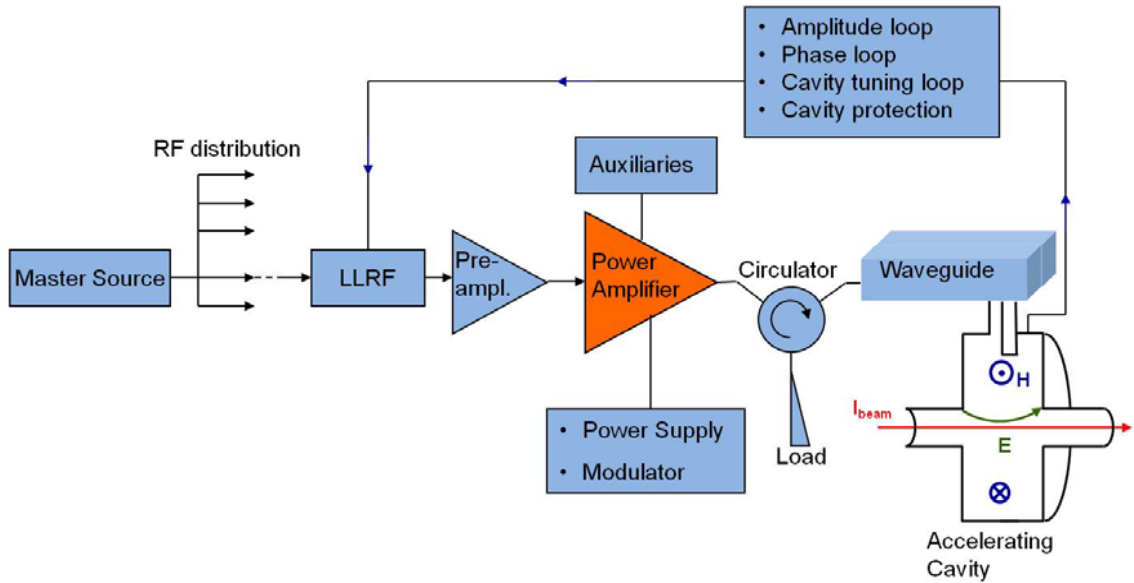

**Fig. 1:** RF transmitter for an accelerating cavity

## 1.2  Example: the ESRF transmitters

The 1.3 MW 352.2 MHz super klystron from Thales Electron Devices (TED, Vélizy/France) in Fig. 2(a) was initially developed for the CERN/LEP ring. Similar klystrons are or were manufactured by other tube manufacturers. Four transmitters using this type of klystron were implemented at ESRF in the 1990s: one to deliver 600 kW in 10 Hz pulses to the booster cavities, and three others to power the storage ring cavities in a continuous wave (CW). As shown in Fig. 2(b), the power from these klystrons needs to be split to feed two to four cavities with several hundred kilowatts each.

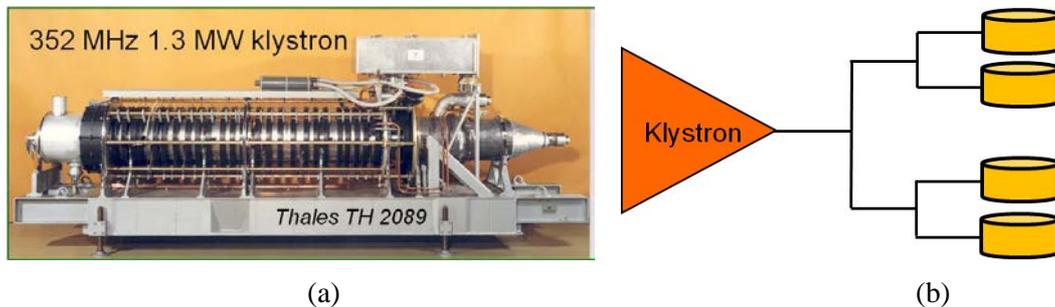

(a)  (b)

**Fig. 2:** (a) 1.3 MW klystron from Thales Electron Devices (TED, Vélizy/France) used at ESRF. DC to RF efficiency $\eta_{max} = 62\%$, gain $G_{max} = 42$ dB. (b) Power splitting to provide several hundreds of kilowatts of RF to a number of ESRF cavities.

These klystron tubes have a high gain and a good efficiency. They are fed from 100 kV, 20 A DC high voltage power supplies with a sophisticated crowbar protection. In modern systems, fast insulated-gate bipolar transistor (IGBT) switched power supplies are used. Klystrons require many auxiliary power supplies for the modulation anode, the filament, and the focusing coils. Due to the high voltage, the electrons generate a substantial level of high energy X-rays when hitting the collector, which necessitates a sophisticated lead shielding hutch.

Out of the original three manufacturers, only one still produces the ESRF klystron shown in Fig. 2(a), and in the early 2000s there remained only a few customers. Seeing the potential risk of obsolescence of these klystrons, it was necessary to become prepared for an alternative RF power source in order to safeguard the operation of the facility. Following SOLEIL's great success with the development and operation of 180 kW SSAs [1] it was decided to implement an initial series of seven SSAs in the frame of the ESRF upgrade phase I over the years 2009 to 2015. Figure 3 shows one of the seven new 150 kW SSAs at ESRF. Four SSAs have replaced the booster klystron transmitter, and each of the three remaining SSAs feeds one new Higher Order Mode (HOM) damped cavity in the ESRF storage ring [5, 6].

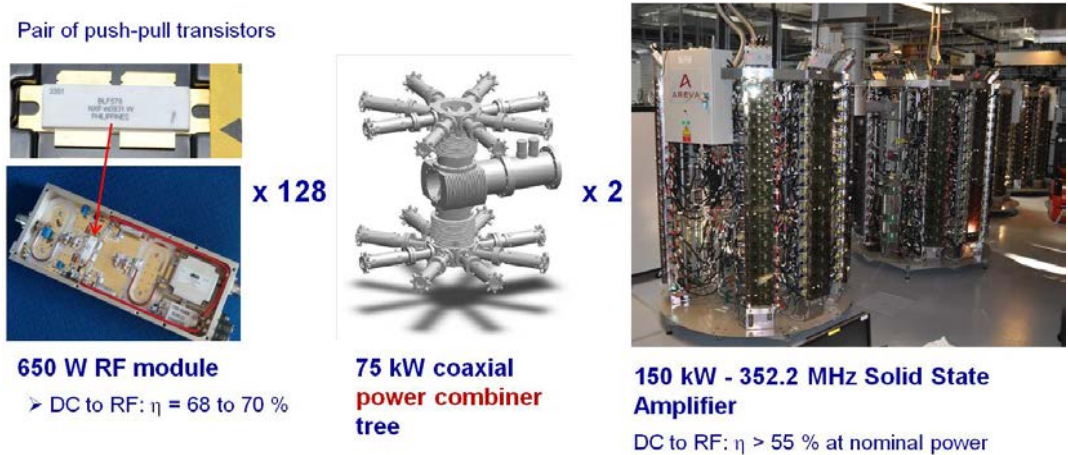

**Fig. 3:** One of the 150 kW–352.2 MHz solid state amplifiers obtained by combining the power of 256 RF transistor amplifier modules each delivering up to 650 W (seven such SSAs are in operation at ESRF, provided by ELTA/AREVA following a transfer of technology from SOLEIL). DC to RF power conversion efficiencies are denoted $\eta$.

An SSA combines the power from as many transistor modules as needed to provide the required nominal output power for one cavity with a given safety margin: 256 modules in the example shown in Fig. 3. The architecture and design of SSAs will be described in detail in Section 2 and a flavour of their performance will be given.

### 1.3 Radio frequency power sources for accelerating cavities – a brief history

This section concludes with a brief history and overview of the RF power sources used to feed accelerating cavities [7]. The graph in Fig. 4 shows the technologies used in various accelerator applications as function of the operating frequency and the required unitary power level.

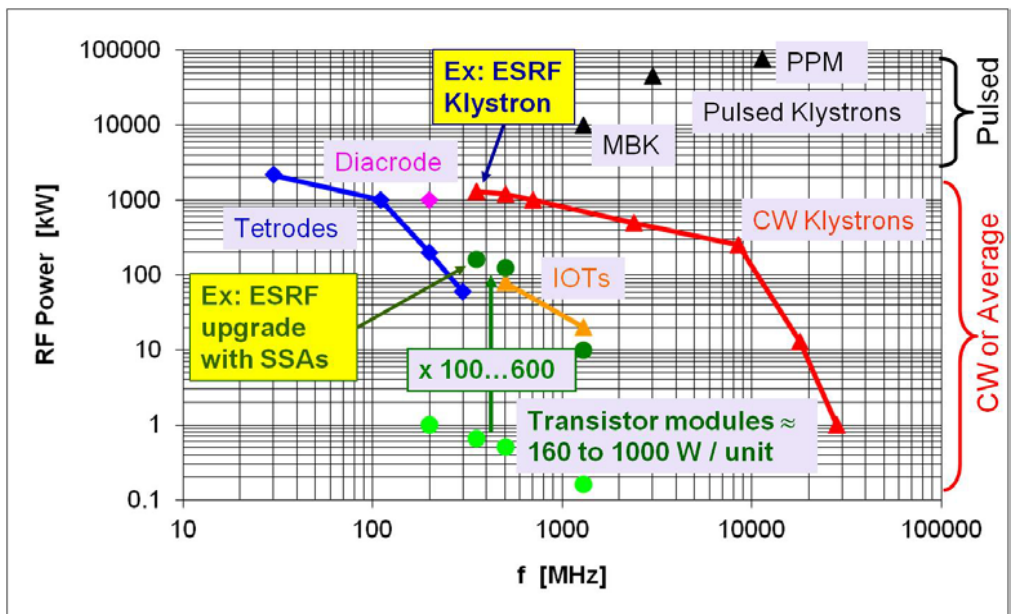

**Fig. 4:** Typical RF sources used in accelerators

Electronic vacuum tubes like triodes and tetrodes have existed since the early twentieth century. Due to their finite electron drift time they are basically limited to frequencies below 1 GHz. Tetrodes are still in use today for broadcast and accelerator applications to generate from kilowatts at 1 GHz up to hundreds of kilowatts below 100 MHz. Note that a small 3.5–5 GHz triode delivering 2 kW pulses exists for radar applications. With its optimized geometry the diacrode developed by Thales can deliver 1 MW at 200 MHz.

In the 1940s and 1950s tubes were developed for high frequency applications that benefit from the electron drift time. They are still widely in use, providing high RF power at high frequencies:

- klystrons: 0.3 GHz–10 GHz, delivering from 10 kW to 1.3 MW in CW and 45 MW in pulsed mode, with applications in TV transmitters, accelerators, and radar;
- IOTs (Inductive Output Tubes or klystrodes): a high efficiency mixture of a klystron and a triode, providing typically 90 kW at 500 MHz to 20 kW at 1.3 GHz, initially developed for SDI (Strategic Defence Initiative) in 1986, with current applications in TV and recently in several accelerators;
- travelling wave tubes (TWT): 0.3 GHz to 50 GHZ, broadband, highly efficient, with applications in satellite and aviation transponders;
- magnetrons: 1 GHz to 10 GHz, narrow band, mostly oscillators, highly efficient for applications in radar and microwave ovens;
- gyrotron oscillators: high power millimetre waves, 30 GHz to 150 GHz, delivering typically 0.5–1 MW pulses of several seconds duration, still subject to much R&D, used for plasma heating for fusion and military applications.

In the 1950s and 1960s the invention and spread of transistor technology also opened the way for many applications in the field of RF:

- bipolar, MOSFET, etc. transistors: delivering several tens of watts up to frequencies of about 1.5 GHz, recently up to 1 kW per unit with sixth generation MOSFETs;
- RF SSAs are increasingly used in broadcast applications, in particular in pulsed mode for digital modulation: 10 kW–20 kW obtained by combining several modules;
- SOLEIL: from 2000 to 2007 SOLEIL pioneered the development of high power 352 MHz MOSFET SSAs for accelerators: 40 kW for their booster, then $2 \times 180$ kW for their storage ring;
- ESRF: recent commissioning of seven 150 kW SSAs, delivered by ELTA/AREVA following a technology transfer from SOLEIL;
- increasing numbers of accelerator labs use, develop, or consider solid state technology for RF amplification, e.g. 1.3 GHz/10 kW SSAs from Bruker at ELBE/Rossendorf, 500 MHz SSAs for LNLS and Sesame designed by SOLEIL, etc.

## RF solid state amplifier

In Fig. 5 is shown an RF SSA with its main components. The RF power from the drive amplifier system is split (2) and distributed to *n* amplifier modules (1) the outputs of which are recombined (3) and coupled to the RF waveguide feeder. A power converter system (4) supplies the amplifier modules with DC power. Components (1) to (4) are described in detail below.

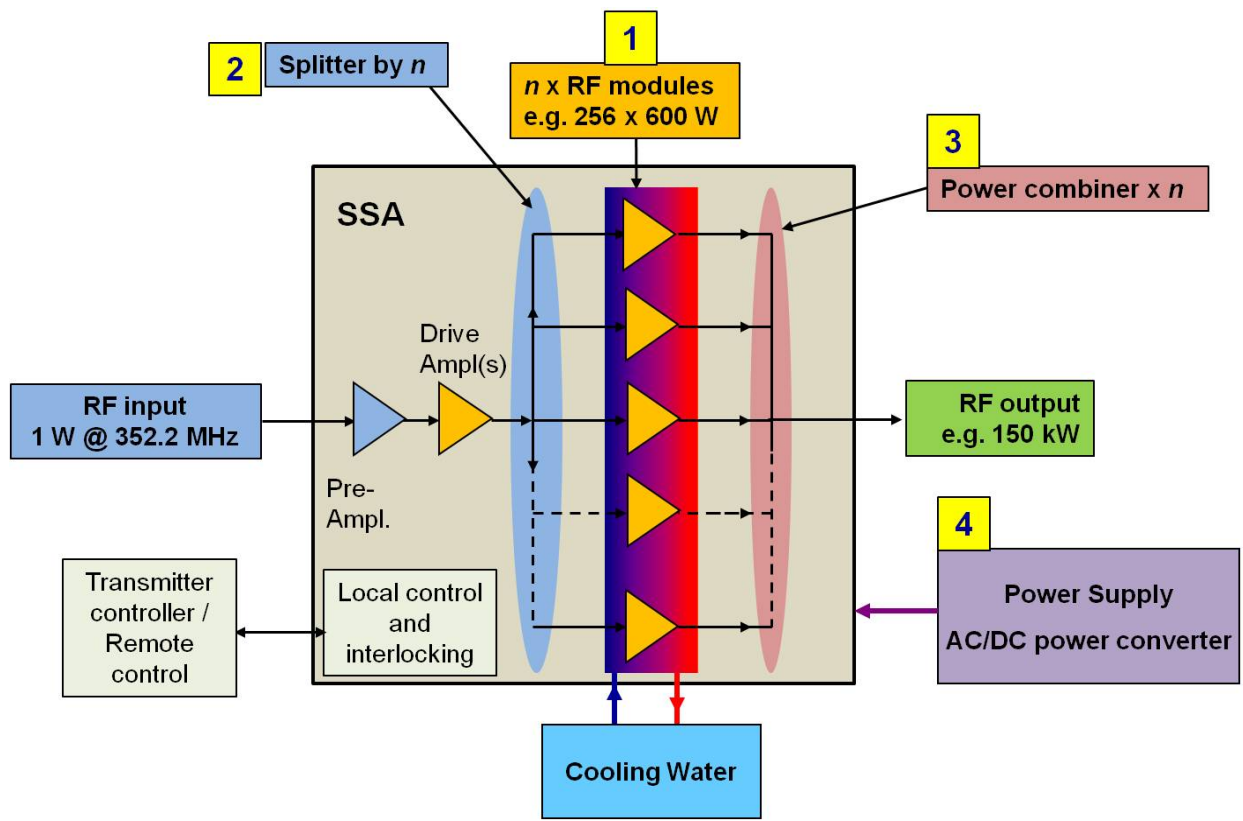

**Fig. 5:** Main components of an RF SSA. The example shown is an ESRF 352.2 MHz–150 kW amplifier

## 2.1 RF amplifier module

### 2.1.1 Transistor

Thanks to the latest developments in high power transistors such as the laterally diffused metal oxide semiconductor field effect transistor (LDMOS FET) and, in particular, the latest sixth generation devices with 50 V drain polarization, it is now possible to generate several 100 W up to 1 kW in CW with a single RF module, for frequencies up to about 1 GHz. Such transistors are, for instance, manufactured by NXP (Eindhoven/Netherlands) and Freescale semiconductor Inc. (Austin Texas/USA).

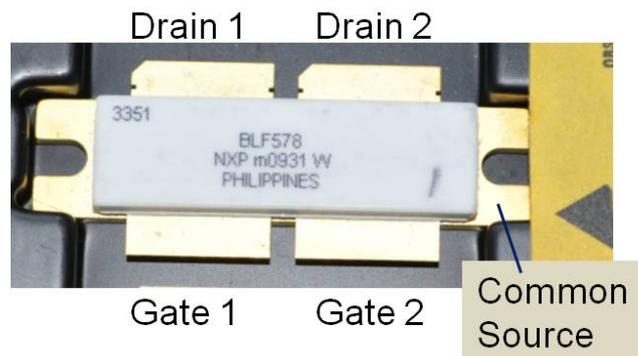

**Fig. 6:** Pair of push-pull transistors

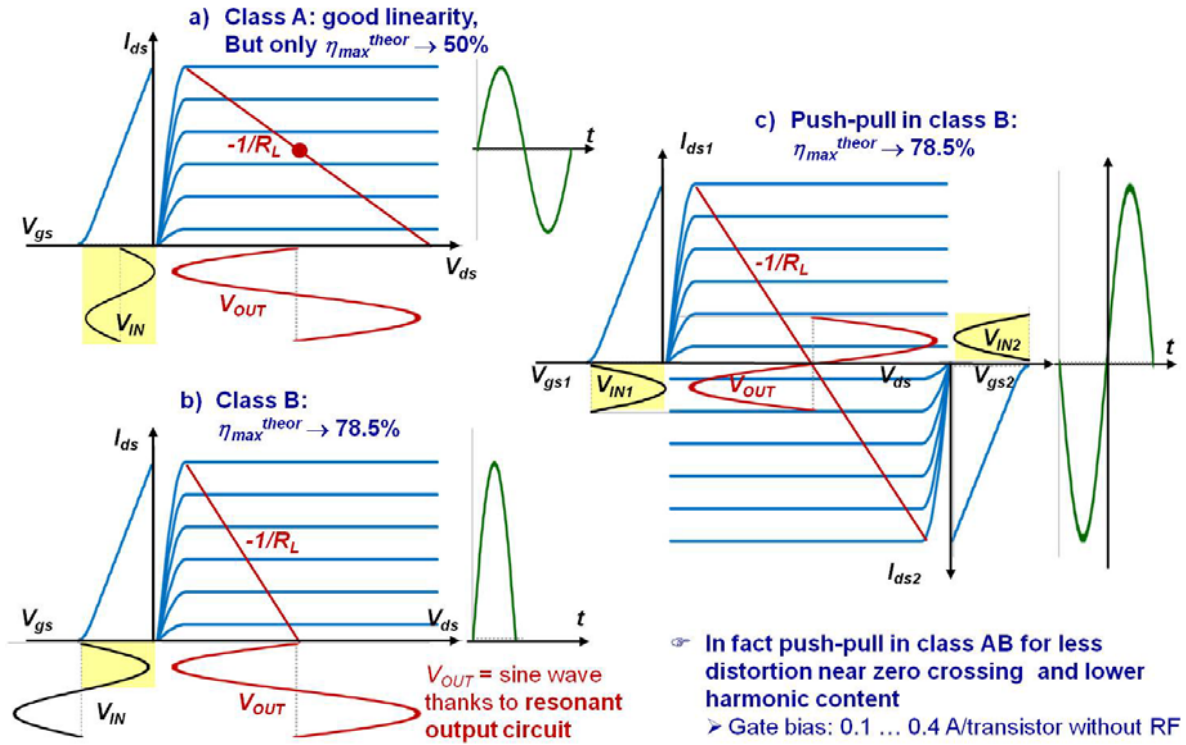

**Fig. 7:** Schematic of MOSFET operation modes. (a) Class A; (b) class B; (c) class B with two transistors in push-pull. In (a, b) drain-source current $I_{ds}$ is given as a function of gate-source voltage $V_{gs}$ on the left sides and as a function of the drain-source voltage $V_{ds}$ on the right sides of the figures; In (c) drain-source currents $I_{ds1}$ and $I_{ds2}$ of both push-pull transistors are given as function of gate source voltages $V_{gs1}$ and $V_{gs2}$ on the left and right sides of the graphs, respectively, and as a function of the respective drain-source voltages $V_{ds1}$ and $V_{ds2}$ with $V_{gs1}$ and $V_{gs2}$ as a parameters in the centre of the graph. The red lines are the working lines of the load resistance $R_L$.

The ELTA/SOLEIL SSA implemented at ESRF uses the BLF 578 transistor from NXP in Fig. 6, which in fact contains two push-pull transistors with a common source, which are operated in class AB. As explained schematically in Fig. 7, this allows optimizing the DC to RF conversion efficiency while keeping the distortion and thereby the harmonic content low: the odd characteristic $I_{ds}(-V_{gs}) = -I_{ds}(V_{gs})$ minimizes the second harmonic H2 and higher even harmonics. As shown in Fig. 7, the working point for an amplifier in class A gives a good linearity as the complete input sine wave $V_{in}$ is folded on the working line $-1/R_L$. But in class A the amplifier always consumes power, even without RF power. At best, the efficiency can approach 50% at full output power. In class B, the amplifier only consumes power when it is driven by RF. This provides a much better efficiency, approaching 78.5% for an ideal transistor. However, only one half of a sine wave is amplified and the signal must be filtered by a narrowband resonator in order to recover the RF signal: it is still polluted by a high harmonic content. Operating two transistors in push-pull in class B allows high efficiency to be maintained, and strongly reducing at least the even harmonics. In fact, the RF modules of the high power SSAs described in this paper are operated in push-pull in class AB with a slight gate bias to further linearize the amplifier characteristics and thereby minimize the harmonic distortions.

### 2.1.2 RF circuit

As shown in Fig. 8, the RF modules are fed with the unbalanced (unsymmetrical) RF signal from an incoming coaxial cable, which is transformed by a balun circuit into two RF signals that are 180° out of phase and drive the pair of transistors. The matching circuit, together with the bias circuit, determine the class AB working point to reach about 1 dB gain compression and a maximum efficiency of about

68–70% at a nominal output RF power of 650 W for the module shown in Fig. 8(c). Each RF module is protected by a circulator with a 1200 W load against reverse power, and can therefore withstand operating with full reflection. The small amount of power leaking backwards through the circulator slightly affects the gain; however, the impact is negligible when operating the device under normal conditions with up to 30% power reflection. The seven ESRF SSAs have been designed such that no power circulator is needed at an output of 150 kW after combining all of the RF modules.

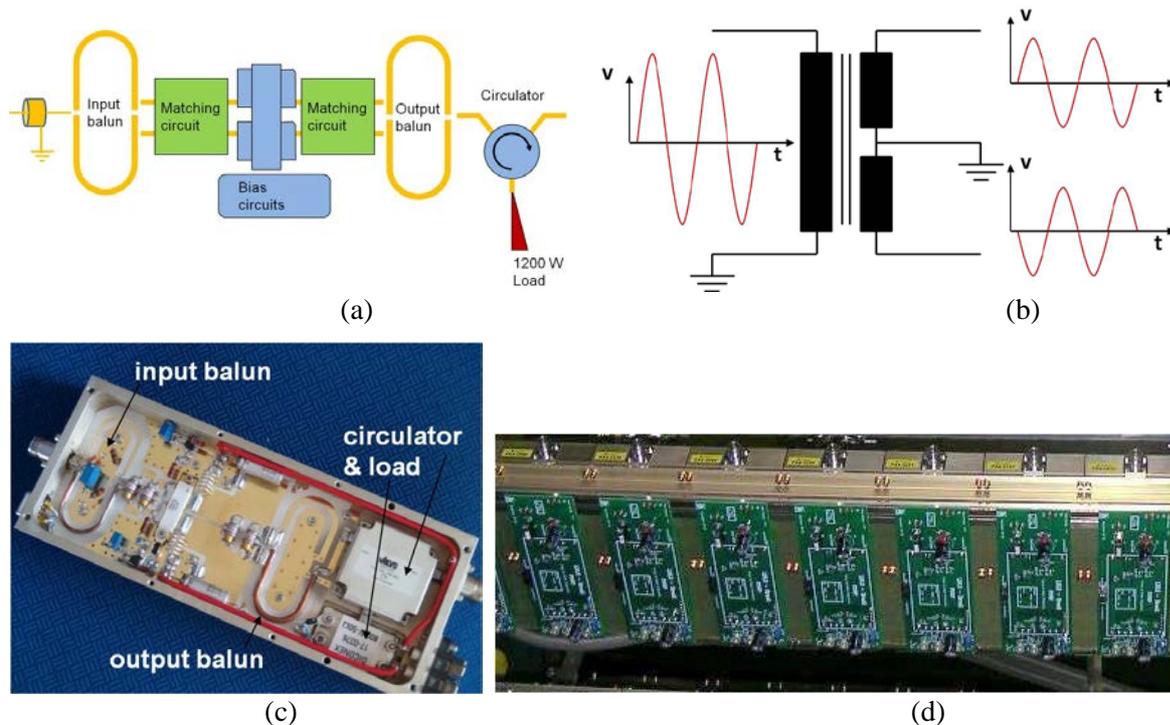

**Fig. 8:** RF circuit. (a) Schematic of the RF circuit comprising input and output balun transformers, matching circuits, a DC bias circuit, and a circulator with a 1200 W load to protect each transistor individually against excessive reverse power. (b) Input balun for the transformation of the unbalanced signal from the coaxial input into a balanced push-pull signal that drives the pair of push-pull transistors. (c) A 650 W ELTA/SOLEIL RF module as installed at ESRF. (d) Water-cooled plate supporting 16 RF modules on the rear side and 32 of the 280 $V_{DC}$/50 $V_{DC}$ converter units (two per RF module) on the front side to feed each module with up to 1.2 kW DC power.

Each RF module is integrated in an individual shielded case. Sensors monitor the temperatures of the transistor socket and the load. The drain current of each pair of transistors is also monitored. These parameters allow the identification of damaged modules during operation.

Note that the input and output balun transformers in Fig. 8(c) are built with hand-soldered coaxial lines.

For the SSA under development at ESRF the fully planar 700 W RF module in Fig. 9 has been designed, and a series of these has been produced to equip a 75 kW prototype SSA. Hand soldering has been almost completely avoided by implementing a planar balun concept from Motorola, by replacing the RF drain chokes with planar quarter-wave transmission lines and minimizing the number of components, all of them surface-mounted devices (SMD) and prone for automated manufacturing. All of these measures allow cost-effective series production of the RF modules.

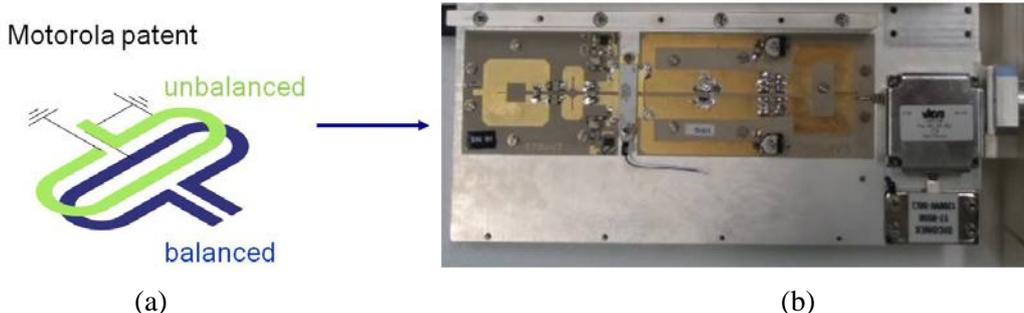

(a) (b)

**Fig. 9:** Development of a fully planar 700 W RF module at ESRF suited for cost-effective mass production. (a) Planar balun patented by Motorola; (b) fully planar ESRF module, efficiency 66% and gain 19.5 dB at 700 W output power.

The parameters given in the caption for Fig. 9(b) are averages measured over the first series of 66 RF modules fabricated with automated SMD techniques. With a gain dispersion of 19.8 ± 0.6 dB and a phase dispersion of ±6°, a good combining efficiency is expected. Many factors contribute to these dispersions, such as the transistor dispersion and other components' parameters, as well as geometrical tolerances. The latter, however, are minimized by the automated pick-and-place technique. Apart from the gate bias, no other parameter has been tweaked.

There is still room for improvement in the achievable efficiency, which is still lower by 3% as compared to the ELTA/SOLEIL RF module. R&D is ongoing in the frame of a collaboration between ESRF and Uppsala University for the optimization of the circuit board.

Note that, as shown in Fig. 5, a few RF amplifier modules are required to amplify the low-level input RF power before splitting it again and distributing it to the inputs of all of the RF amplifier modules. In the ELTA/SOLEIL design, one pre-amplifier feeds four RF modules, each of which then feeds 64 RF power modules.

## 2.2 Power splitters

Power splitters are key components, designed to distribute the RF power from the drive amplifiers to the individual RF modules with minimum amplitude and phase dispersion. At 352 MHz a matched splitter distributing the drive power without reflections can be built as shown in Fig. 10. Each quarter-wavelength line transforms the 50 Ω at its output to $Z^2/(50\ \Omega) = n \times 50\ \Omega$ as seen from its input. The parallel connection of these $n$ lines is thus matched to the 50 Ω of the incoming line.

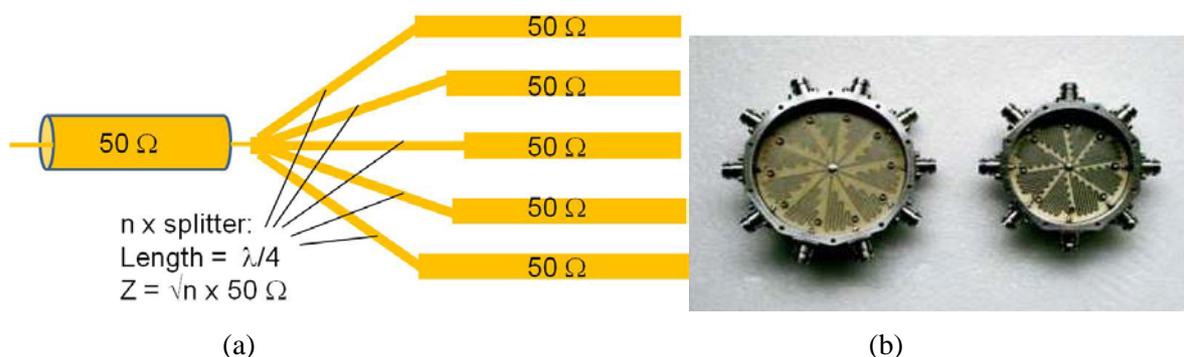

(a) (b)

**Fig. 10:** Matched RF stripline splitters for $n$ outputs using quarter-wavelength ($\lambda/4$) transformers. (a) Schematic representation; (b) SOLEIL 10 and 8 line splitters (lids removed) [1, 2].

The Wilkinson splitter in Fig. 11 works in the same manner as the simple quarter-wavelength splitter shown in Fig 10. The additional 50 Ω resistors are not seen by the incoming split signals as they have the same amplitude and phase (common mode). However, differential signals are absorbed by these loads, thereby decoupling the outgoing arms from each other. With a Wilkinson splitter, the connected amplifier inputs are therefore decoupled. Such Wilkinson splitters are implemented in the SSA in development at ESRF.

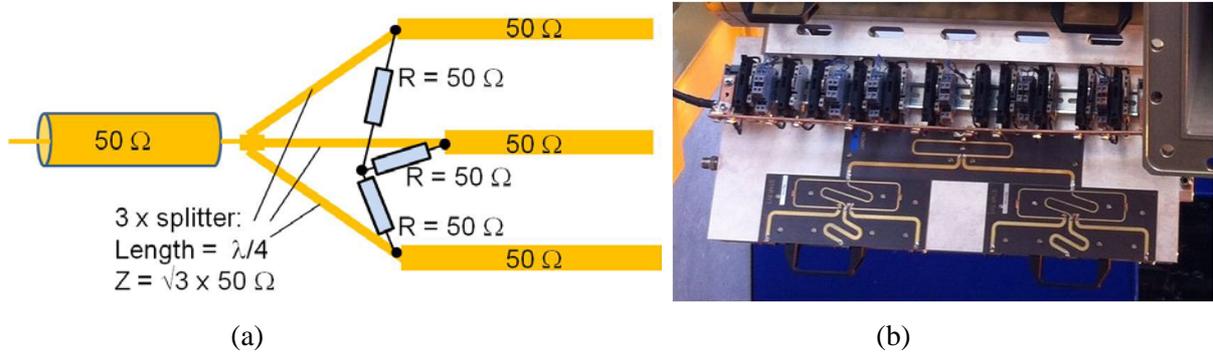

(a)            (b)

**Fig. 11:** Wilkinson splitters: resistors absorb differential signals without perturbing the common mode, thereby decoupling the connected outputs from each other. (a) Schematic representation; (b) implementation on the SSA under development at ESRF.

## 2.3 Power combiner

SSAs are often built with several stages of power combiners that use quarter-wavelength transformers like the splitters shown in Fig. 10, except that they are operated in reverse and that the striplines are replaced with transmission lines that are adapted to higher power levels, mostly larger coaxial lines. As for the power splitters, the quarter-wavelength transformers guarantee that the combiner tree is matched in impedance to the connected high power output, provided that all of the inputs are at 50 Ω, which is the case thanks to the circulator/load system at the output of each RF amplifier module. However, as the combiners are non-directional, each individual amplifier does not at all see a matched impedance.

### 2.3.1 *General considerations on power combiners*

Before entering into the technology, let us first address some general characteristics of such power combiners. We consider a power combiner with $n$ inputs in a single stage. One can easily derive the S-parameter matrix, which has the following general form:

$$\begin{bmatrix} b_1 \\ b_2 \\ b_3 \\ \ldots \\ b_n \\ b_{n+1} \end{bmatrix} = \begin{bmatrix} (1-n)/n & 1/n & 1/n & \ldots & 1/n & 1/\sqrt{n} \\ 1/n & (1-n)/n & 1/n & \ldots & 1/n & 1/\sqrt{n} \\ 1/n & 1/n & (1-n)/n & \ldots & 1/n & 1/\sqrt{n} \\ \ldots & \ldots & \ldots & \ldots & \ldots & \ldots \\ 1/n & 1/n & 1/n & \ldots & (1-n)/n & 1/\sqrt{n} \\ 1/\sqrt{n} & 1/\sqrt{n} & 1/\sqrt{n} & \ldots & 1/\sqrt{n} & 0 \end{bmatrix} \times \begin{bmatrix} a_1 \\ a_2 \\ a_3 \\ \ldots \\ a_n \\ a_{n+1} \end{bmatrix}, \quad (1)$$

where

$a_1 \ldots a_n$ are the incident wave amplitudes at the input arms (coming from the amplifier modules);

$b_1 \ldots b_n$ are the reverse wave amplitudes at the input arms (coming back to the amplifier modules);

$b_{n+1}$ is the wave amplitude leaving the output arm of the combiner;

$a_{n+1}$ is the reverse wave amplitude incident on the output arm (e.g. the reflection from the cavity).

Note that wave amplitudes $a_i$, $b_i$ are measured in $\sqrt{W}$ and are generally complex, their arguments representing the respective phase of the wave at port $i$:

$|a_i|^2$ is the power of the wave incident on port $i$, where $a_i = |a_i|\, e^{j\varnothing_i}$, and $\varnothing_i$ is the phase of the incident wave $i$;

$|b_i|^2$ is the power of the wave coming from port $i$, where $b_i = |b_i|\, e^{j\varnothing'_i}$, and $\varnothing'_i$ is the phase of the reverse wave $i$.

The different terms in the S-matrix are also generally complex, their arguments depending on the exact geometry of the combiner. However, pertaining to the necessary symmetry of a single stage power combiner, all of the S-parameters $(1 - n)/n$ have the same arguments, as do all of $1/n$ and all of $1/\sqrt{n}$. The phases of the $(1 - n)/n$ and the $1/n$ are 180° apart. But the conclusions given below hold even if we neglect the additional phase terms.

a) Each connected RF module sees individually a strong mismatch with a reflection coefficient $(1 - n)/n$ that approaches $-1$ for a large $n$, corresponding to nearly full reflection.

b) The individual inputs for the amplifier modules are coupled to each other with a coefficient $1/n$ showing that this kind of combiner has no directivity at all.

c) However, *when all of the input waves are equal in amplitude and phase,* say with an amplitude $a_0$, their signals interfere perfectly well in the following way.

   i) For any $i, j \leq n$,

   $$a_i = a_0 \Rightarrow b_j = a_j\,(1 - n)/n + \Sigma\,[a_{i \neq k, \leq n}\,(1/n)] = [(1 - n)/n + (n - 1) \times 1/n]\,a_0 = 0\,,\qquad(2)$$

   meaning that *any of the RF amplifier modules operates in matched condition* without any reflection. The strong reflection of its own signal is exactly compensated by the sum of the signals coupled from all other amplifier modules that have the opposite phase,

   ii) $b_{n+1} = \Sigma\,[a_{i, \leq n}\,(1/\sqrt{n})] = [n/\sqrt{n}]\,a_0 \Rightarrow |b_{n+1}|^2 = n\,|a_0|^2\,,\qquad(3)$

   meaning that, neglecting imperfections, *the power from all n inputs combines to 100% in the output arm* of the combiner, as expected.

So, the power adds up correctly at the output of a combiner only by constructive interference, i.e. if and only if the signals at their inputs have the same amplitude and phase. Any differential mode power is distributed as reverse power to the individual input arms, i.e. to the output stages of the RF amplifier modules, where it is absorbed by the circulator loads. So, good control of a flat amplitude and phase distribution are crucial for obtaining a good combining efficiency.

d) *If just one RF amplifier on arm i is switched off*, it will receive the reverse power coupled from all of the other arms without compensation by its own reflection:

$$b_i = \Sigma\,[a_{j \neq i, \leq n}\,(1/n)] = [(n - 1)/n]\,a_0\,,\qquad(4)$$

which approaches $a_0$ for a large $n$. The consequences are as follows.

   i) When one RF module is not providing power to the combiner, it receives from all the other modules a reverse power that is equivalent to the power of one module.

   ii) When one RF module is not providing any power to the combiner, the output power is lacking the equivalent power from two modules: the power from the missing module plus the power absorbed in the circulator load of the missing module.

   iii) When the SSA output is working on a mismatch with a reflection coefficient $r = |r|\,e^{j\varnothing}$, the reverse wave of amplitude $a_{n+1} = r\,b_{n+1}$ couples back to each module, including the unpowered RF module, as much as

   $$a_{n+1}/\sqrt{n} = r\,b_{n+1}/\sqrt{n} \approx r\,(n \times a_0/\sqrt{n})/\sqrt{n} = r\,a_0\,.\qquad(5)$$

In total, an unpowered amplifier module on an arm *i* will see reverse power corresponding to the following reverse wave amplitude:

$$b_i = [(n-1)/n + r]\, a_0 \approx [1 + |r|\, e^{j\emptyset}]\, a_0 \Rightarrow |b_i|^2 \approx |1 + |r|\, e^{j\emptyset}|^2\, |a_0|^2, \qquad (6)$$

with the worst case

$$\emptyset = 0 \Rightarrow |b_i|^2 \approx (1+|r|)^2\, |a_0|^2. \qquad (7)$$

For full reflection $|r| = 1$, a non-powered RF module can therefore see reverse power as high as four times the nominal power of one module.

For the ELTA SSA, ESRF had specified a maximum reflection of one third of the incident power. Given the nominal module power of 630 W, this means that in the worst case as much power as

$$P_{\text{reverse}}^{\max} = |b_i|^2 \approx (1 + 1/\sqrt{3})^2 \times 630\ \text{W} = 1570\ \text{W} \qquad (8)$$

can come back to the circulator and then into the load of an unpowered RF module. Such values were indeed measured on the real SSA shown in Fig. 3. The way to avoid overloading the 1200 W circulator loads of the RF modules is described in Section 2.3.2.

### 2.3.2 *Coaxial combiner*

In most of the existing high power SSAs the power from the individual RF amplifier modules is combined by means of several stages of cascaded coaxial combiners. The SOLEIL SSAs and the 150 kW SSAs in operation at ESRF are literally built around coaxial combiner trees that become amplifier 'towers' in the final assembly. As shown in Fig. 3, the power from two such 75 kW towers is further combined to feed up to 150 kW into an outgoing WR2300 waveguide. The size of the coaxial combiners depends on the power level as, for example, shown in Fig. 3:

- first stage: 630 W × 8 by means of EIA[2] 1 5/8″ coaxial combiners yielding 5 kW;
- second stage: 5 kW × 8 by means of EIA 6 1/8″ coaxial combiners yielding 40 kW;
- third stage: 40 kW × 2 by means of EIA 6 1/8″ coaxial combiners yielding 80 kW;
- fourth stage: 80 kW × 2 by means of an EIA 9 3/16″ coaxial combiner yielding 160 kW (maximum power, the nominal value being 150 kW).

*Dimensioning of the circulator loads and fine tuning of the coaxial combiner*

Initially, the circulator loads of the 150 kW SSAs for ESRF were dimensioned so as to withstand full module power up to 650 W maximum in full reflection at any phase. This was met by selecting a corresponding circulator and an 800 W load. The SSAs were also tested successfully at full power with up to 30% power reflection at any phase. The SSAs had furthermore been designed with sufficient power reserve to benefit from the intrinsic redundancy of many combined RF amplifier modules, and full output power could easily be fed into a matched load with up to six missing RF modules! However, when testing the first SSA at full power with 30% reflection and some of the modules switched off, the output circuits and the loads of the unpowered RF modules were damaged, while the phase of the mismatch was varied. In fact, operating conditions as described in Eqs. (6–8) lead to an overloading of the circulator loads of the non-powered RF modules. As much as 1500–1700 W was measured depending on the location in the combiner, confirming experimentally the 1570 W predicted in Eq. (8).

Fortunately, the coaxial combiner in the 150 kW ESRF SSA has multiple stages. Its S-parameter matrix does not have the exact symmetry of the single-stage combiner in Eq. (1). When all *n* of the RF modules feed the combiner with equal amplitudes and phases, their power still combines perfectly into the output port $n + 1$. Depending on the length of the output arms of the combiners, for instance in the first stage as shown in Fig. 12, the reverse power coming to an RF module from its seven immediate neighbouring RF modules interferes more or less constructively or destructively with the power coming from the other RF modules [8]. The tests had revealed that this interference was close to maximum when the load was burnt. As suggested by the experts from SOLEIL [8] and shown in Fig. 12, this interference could be made destructive by inserting 170 mm long delay lines at all of the outputs of all of the first combiner stages. The measurements shown in Fig. 12(b) show that the load power of a switched-off module then remained below 1200 W when operating the SSA at the specified maximum mismatch and varying the phase of the reflected wave. To complement the modification shown in Fig. 12, the 800 W circulator loads were replaced with 1200 W loads, as already shown in Fig. 8.

---
[2] EIA: US American Electronic Industries Alliance standard

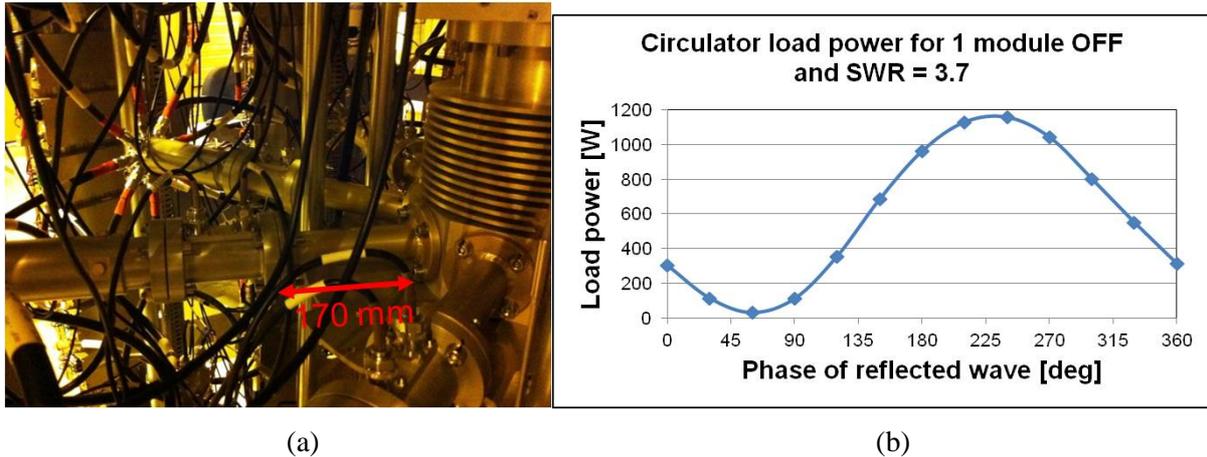

(a)          (b)

**Fig. 12:** Minimization of the circulator load power of a switched-off RF module when the 150 kW SSA operates on a mismatch corresponding to 30% power reflection. (a) Insertion of a 170 mm delay line at the output of the first combiner stage; (b) measured load power is reduced from maximum 1700 W down to 1200 W.

The modifications described above were implemented for the three SSAs that are operated in continuous wave (CW) on the ESRF storage ring. This was not necessary for the ESRF booster, which is run in pulsed mode and where the relevant average load power remains well below 800 W.

### 2.3.3 Cavity combiner developed at ESRF

A compact single-stage cavity combiner is under development at ESRF. As shown in Fig. 13, its $E_{010}$ resonance has a homogeneous azimuthal and longitudinal field distribution. For the 352.2 MHz ESRF application, the outer cylindrical wall is equipped with six vertical rows of 22 input loops distributed around the circumference. In total 132 transistor modules like that shown in Fig. 9, delivering up to 700 W each, will be connected to these coupling loops. The total output power is expected to be between 80 kW and 90 kW.

The coupling $\beta_{\text{module}}$ must be the same for all of the input loops. It is easily adjusted by means of the loop size (typically a few square centimetres). The coupling of the output waveguide $\beta_{\text{waveguide}}$ is mainly determined by the size of the capacitive disc that couples to the electrical field of the $E_{010}$ resonance. To obtain matched conditions and an optimum combining efficiency for a number $n$ of RF modules, the coupling factors must be set according to:

$$\beta_{\text{waveguide}} \approx n \times \beta_{\text{module}} \gg 1 \ . \tag{9}$$

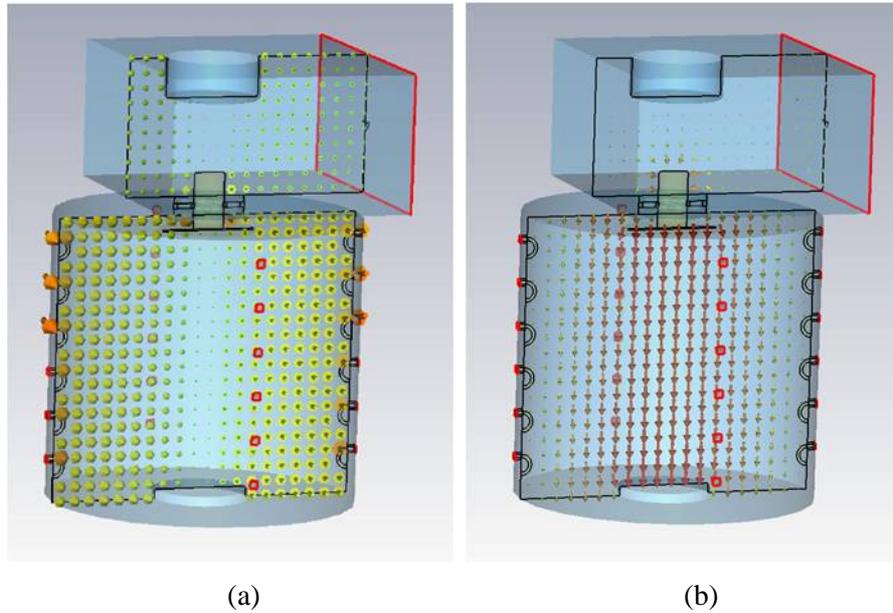

(a)                  (b)

**Fig. 13:** Strongly loaded $E_{010}$ resonance of a cylindrical resonator to combine the power from 132 RF modules in a single stage. (a) H field: homogeneous magnetic coupling to 132 input loops on the cylindrical wall; (b) E field: strong adjustable capacitive coupling to the output waveguide.

As shown in Fig. 13 and implemented on the ESRF prototype in Fig. 14, the piston attached to the top of the waveguide can be moved up and down, and the back short-circuit plane can be moved back and forth. This allows adjustment of the coupling factor $\beta_{module}$ in a range of about 1:3 and thereby matching the SSA for a variable number of connected modules. The nominal power of the SSA can therefore be easily adapted to changing operating conditions by simply removing a number of RF modules and re-adapting the waveguide coupling accordingly.

Figure 14 shows the prototype ESRF amplifier with cavity combiner. The water-cooled 'wings' that support six RF modules each constitute a section of the cavity wall with built-on coupling loops. The RF modules are hence directly flanged to the combiner and there is no need for coaxial RF power cables. The RF splitters for the distribution of the RF drive power as well as the DC power distribution are fixed onto the rear sides of the wings. The prototype in Fig. 14 with only three active wings and 19 blind flanges delivered as much as 12.4 kW of RF power with a DC-to-RF conversion efficiency of 63%.

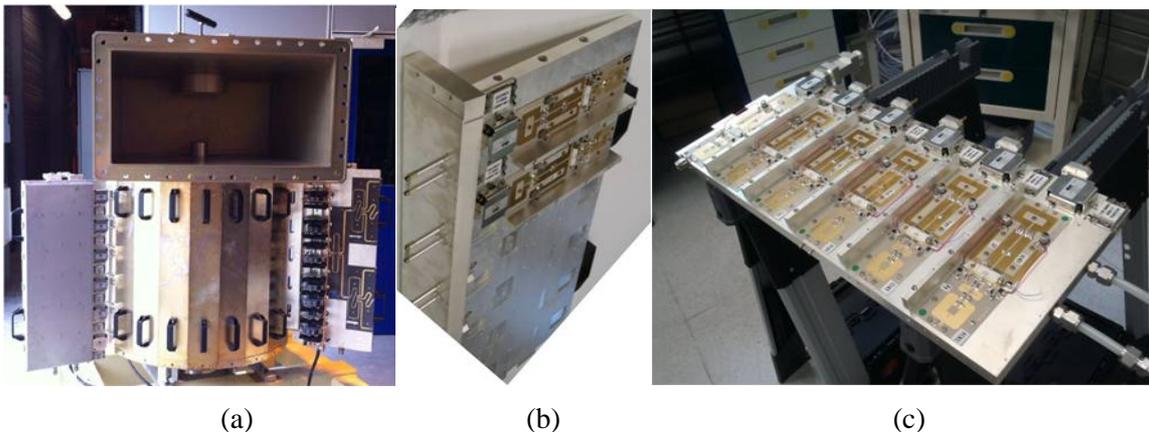

(a)                  (b)                  (c)

**Fig. 14:** Prototype with three wings and a total of 18 700 W RF modules, successfully tested at 12.4 kW with a DC-to-RF efficiency of 63%. (a) Cavity combiner with WR2300 output waveguide; (b) direct coupling of RF modules; (c) water-cooled wing with six RF amplifier modules.

The extension of the prototype with a total of 22 active wings is close to completion and will be ready for power tests in 2015. The achievable output power is estimated to be between 80 kW and 90 kW, which is even slightly above the power obtained from one ELTA/SOLEIL coaxial tower. At the same time, the cavity combiner is more compact by far than an equivalent coaxial combiner.

## 2.4 Power supply

As shown in Fig. 8 the ELTA/SOLEIL SSAs are equipped with two 280 $V_{DC}$-to-50 $V_{DC}$ converters per RF module. Each of the three 150 kW SSAs in operation on the ESRF storage ring is fed with one 300 kW/400 $V_{AC}$-to-280 $V_{DC}$ converter, which was developed by the ESRF Power Supply Group. As may be seen in Fig. 15, a 280 $V_{DC}$-to-RF conversion efficiency of 57% is obtained at 150 kW output power. Beyond nominal power, the efficiency approaches 60%.

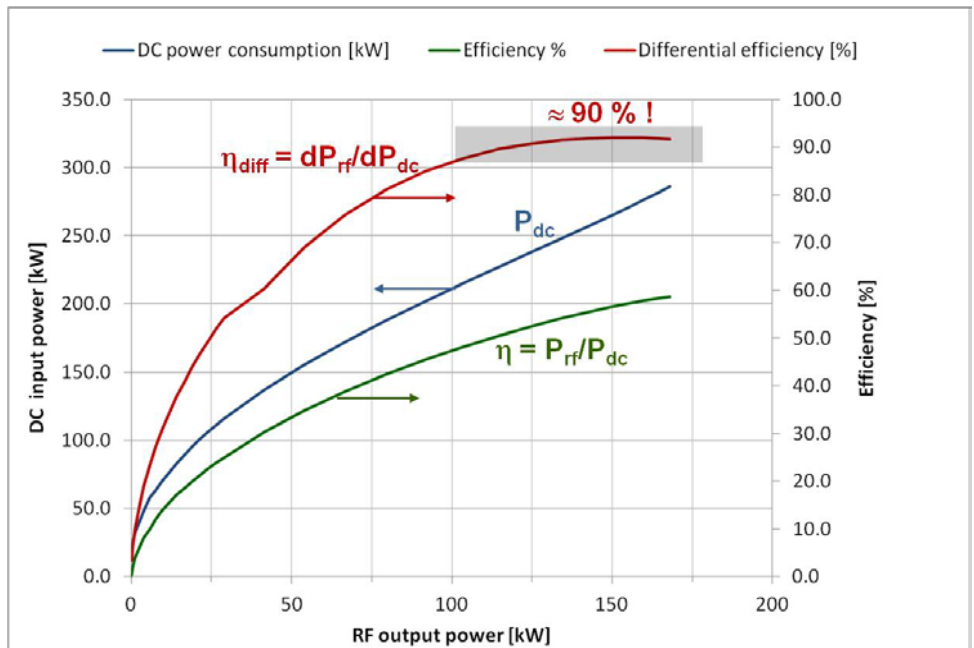

**Fig. 15:** DC-to-RF efficiency and required DC power as a function of the output RF power of the ELTA/SOLEIL SSA implemented at ESRF.

It is worth noting that the efficiency drops fast when operating the amplifier at reduced power, and is only about 47% at 100 kW, i.e. at ⅔ of the nominal power. Note also that in the upper range 100–150 kW, the differential efficiency is about 90%, meaning that, with 1 kW of additional DC power, one obtains as much as 0.9 kW of additional RF power. In other words: in order to make good use of the electrical power, it is important to correctly dimension the RF system and operate the SSAs close to their nominal power. Unfortunately, when designing an RF system, it is generally common practice to slightly over-dimension the available power in order to be able to overcome unforeseen transmission losses and to be prepared for possible power upgrades of the accelerator.

A nice feature of the cavity combiner described in section 2.3.3 is that one can at any time remove a number of RF modules and easily re-adapt the output coupling. This way it is easy to bring down the nominal power of an over-dimensioned SSA close to required operation level and recover a good overall efficiency. Reference [9] describes another method for recovering good efficiency at reduced power by modulating the drain voltage of the transistors. However, it is important to check the stability of the amplifier modules over the entire range of the drain bias.

Four 150 kW SSAs feed the accelerating RF cavities of the ESRF booster. Currently, a system of resonant AC and DC power supplies feed the booster magnets with a 10 Hz sine wave. The dipole magnet currents determine the modulation of particle energy $E$ in the booster, as shown in Fig. 16. The

RF voltage $V_{tot}$ performs the bunching of the beam injected from the linac at 200 MeV and accelerates it to 6 GeV where it is extracted to the storage ring. In this process, the largest portion of RF voltage is needed to compensate for the synchrotron radiation losses that scale with $E^4$ and lead to the red curve for the total accelerating voltage $V_{tot}$ in Fig. 16. The total DC power $P_{dc\text{-}tot}$ absorbed by the SSAs to produce the RF power $P_{rf}$ needed to obtain this voltage is oscillating at 10 Hz between about 100 kW and 1000 kW. An anti-flicker capacitor bank of 3.2 F has been installed to prevent this power fluctuation being transmitted to the mains. The residual flicker thus remains well below the legal limit of 0.29%. A good side effect is that an AC power of maximum 400 kW, slightly above the average DC power $P_{dc\text{-}aver}$ in Fig. 16, is drawn from the mains. This constitutes a reduction of nearly a factor 3 in power consumption with respect to the previous klystron transmitter, for which a 10 Hz filtering was not possible at 80 kV supply voltage. The klystron was operated with input RF modulation and constant beam power, most of this power being absorbed in the klystron collector.

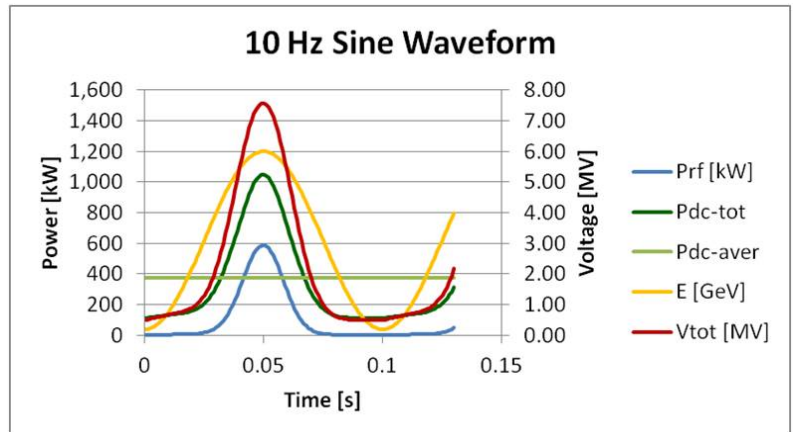

**Fig. 16:** Pulsed RF for the ESRF booster

For the SSA with the cavity combiner under development at ESRF, the RF modules will be fed from 22 power converters, one per wing, with a direct conversion from 400 $V_{AC}$ to 50 $V_{DC}$. They will be installed close to the SSA in order not to carry the high currents over longer distances.

## 2.5 Specification

Having addressed the main ingredients of a high power RF SSA, typical performance data are now reviewed, however not exhaustively, based on the 352.2 MHz–150 kW SSAs delivered by ELTA to ESRF.

- The specified 280 $V_{DC}$-to-RF conversion efficiency was easily met:
    - ✓ $\eta > 57\%$ at $P_{nom} = 150$ kW   (specified: $\eta > 55\%$);
    - ✓ $\eta > 47\%$ at ⅔ $P_{nom} = 100$ kW   (specified: $\eta > 45\%$).
- Gain compression <1 dB at $P_{nom} = 150$ kW.
    - o Figure 17 shows the typical gain curve of one 150 kW SSA at ESRF. The P1 point, at which the gain compression reaches 1 dB, has been set above the nominal output power to limit distortion and safeguard the transistor against drain over-voltage. The P1 point is adjusted by means of the load resistance $R_L$ in Fig. 7.
- Avoid overdrive conditions.
    - o High peak drain voltage can damage the transistor. It is thus crucial to implement an overdrive protection interlock.
- Short pulses (20 µs).

- o RF cavities and their power couplers need to be slowly conditioned to high RF power, starting with cycles of 20 µs short pulses, then increasing the pulse length up to CW operation.
- o Up to 1.3 dB transient gain increase has been measured in pulsed operation, which bears a risk of overdrive. The overdrive protection therefore needs to be adjusted carefully.
– Requested redundancy for reliable operation.
- o All of the points from the specification in terms of power and operating conditions are achieved with up to 2.5% missing RF modules, i.e. with up to six RF modules in fault. As never more than one or two modules fail at the same time, the SSA never stops for output module failures. One can wait until the next programmed maintenance shutdown to exchange faulty modules.
- o The requested redundancy requires some power margin that is paid for with a slightly lower efficiency than that which is ultimately achievable. The power margin thus needs to be dimensioned carefully.
- o Note that if one of the RF power modules used in the drive chain fails, the output power can no longer be maintained and the SSA may trip the accelerator.
– Harmonics:
- ✓ H2 < −36 dBc    (36 dB below carrier at 704.4 MHz);
- ✓ H3 < −50 dBc    (50 dB below carrier at 1056.6 MHz).
– Spurious sidebands/phase noise:
- ✓ Sidebands < −68 dBc at 400 kHz    (DC/DC converter switching, harmless).
- ✓ A substantial improvement as compared with −50 dBc for klystrons from voltage ripples of the HV power supplies at 600 Hz, 900 Hz, 1200 Hz, etc., which are, moreover, much closer to the synchrotron frequency at which the stored beam is extremely sensitive.
– Operation on mismatched load:
- ✓ 150 kW output power with ⅓, i.e. 50 kW, reflection at any phase,
- ✓ 80 kW output power with 100% reflection at any phase,
- ✓ 150 kW output power with 100% reflection at any phase for a duration of up to 20 µs.
- o Note that for high power reflections, due to the limited isolation of the circulators of the RF modules, the residual reverse power reaching the transistor output circuit provokes a gain modulation. Also, the power combiner itself contributes to some gain modulation in the presence of high reflected power. As a result, the specified output power cannot be reached for some phase values of the mismatch, due to the overdrive limitation.
– Reliability:
- ✓ Not more than 0.7% of the installed RF power modules, including their individual DC/DC converters, are allowed to fail within one year.
- o With seven 150 kW SSAs gradually commissioned at ESRF between early 2012 and the end of 2013, not including early failures with less than 1000 hours of operation (debugging), the failure rate is in the range of the specified 0.7% per year. However, the cumulated operating time is still too low to give a statistically sound statement.

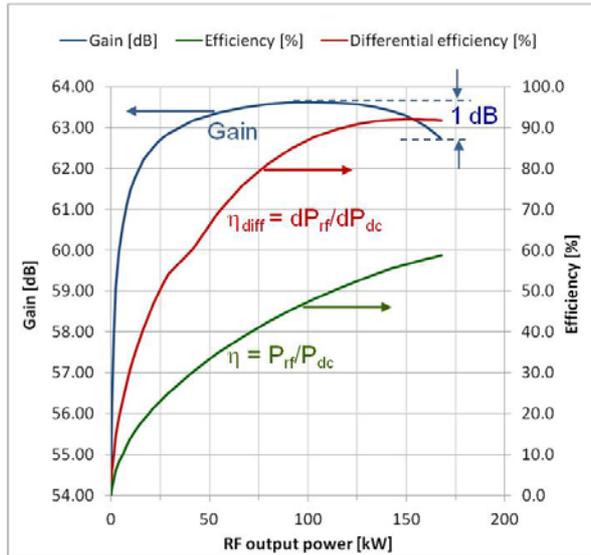

**Fig. 17:** Gain and efficiency of an ESRF 352.2 MHz–150 kW SSA delivered by ELTA

## 2.6 Transient reflections for pulsed cavity conditioning

For the RF conditioning of accelerating cavities they are fed with short RF pulses. The physical explanation for the transient wave amplitudes in Fig. 18(a) is straightforward:

– the leading edge of the incoming RF pulse sees the cavity as a short circuit and is reflected with a factor of −1;

– after the filling time of the cavity ($T_f = Q_o/(\pi f_{rf} (1 + \beta)) = 6.1$ µs), the reflection coefficient approaches the steady-state value $r = (\beta − 1)/(\beta + 1) = 0.58$;

– the negative falling edge of the incoming pulse again sees a short with $r = −1$ and gives rise to a positive transient reflection of amplitude 1 that adds to the 0.53 amplitude of the almost steady reflection: the total transient reflection therefore peaks at more than 1.5 times the incoming pulse amplitude, corresponding to more than twice the input power;

– after the cavity filling time of 6.1 µs, the reflected pulse vanishes;

– The measurement of the incident and reflected power transients shown in Fig. 18(b) confirms this behaviour: the scope traces correspond to the squared amplitude plots shown Fig. 18(a).

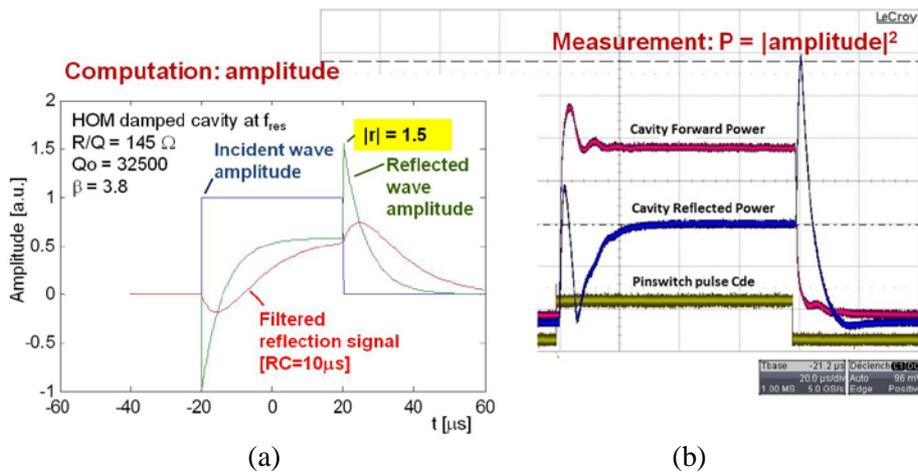

(a)                    (b)

**Fig. 18:** Transient reflections when conditioning an accelerating cavity with short RF pulses. (a) Calculated transient wave amplitudes; (b) measured transient wave power.

The SSAs are protected against reflections exceeding the specified 50 kW as noted above. However, for a duration below 20 µs a reverse power of 150 kW is acceptable. Therefore, in order not to trip the SSAs for pulsed cavity conditioning, the signal triggering the 50 kW interlock was filtered as shown in the red curve in Fig. 18(a). This interlock then only triggers for power exceeding the maximum cavity conditioning power of 80 kW. The SSA is still protected by a second unfiltered fast interlock that triggers if the reflected power transient exceeds 150 kW. Taking into account the power reflection on the falling edge of the pulse, this interlock will trigger if the incident power exceeds the 80 kW needed for cavity conditioning.

## Conclusions – a short comparison between klystrons and solid state amplifiers

To conclude this paper the pros (+) and cons (−) of implementing high power RF solid state amplifiers as an alternative to klystrons are tentatively addressed.

- \+ SSAs do not need a high voltage power supply (50 V instead of 100 kV):
    - \+ consequently there is no need for X-ray shielding;
    - \+ 20 dB less phase noise.
- \+ High modularity/redundancy providing a high reliability in operation:
    - o the SSA is still fully operational if a few RF modules fail, except if a driver module fails.
- − More space required per kW for a SSA than for a klystron:
    - o however, thanks to the modularity, it is easier to precisely match the power of an SSA to the requirements;
    - o the cavity combiner provides a substantial size reduction when compared to a coaxial combiner tree.
- • Durability and obsolescence.
    - o Klystrons: there is no problem with klystrons and tubes in general for as long as a particular model is still manufactured. But it can become very problematic in the event of obsolescence, as the development costs for new tubes are too high for medium-sized labs.
    - o SSAs: transistors generally have a shorter product lifetime. However, there will always be comparable or even better transistors on the market. Nevertheless, operating SSAs commits the user to carefully follow the transistor market and react quickly enough to develop RF modules with new transistors that fit into the existing SSAs.
- \+ Easy maintenance of SSAs if sufficient spare parts are available.
- • Investment costs:
    - − SSAs have still a higher price per kW than comparable tube solutions.
    - \+ But SSA technology is progressing. A significant cost reduction is for instance expected from the possible mass production of fully planar RF modules as the ones developed at ESRF (see Section 2.1.2). Also the use of compact cavity combiners will reduce the fabrication costs as compared to building up large coaxial combiner trees.
    - \+ Prices for SSA components should sink while prices for klystrons have strongly increased over the last decades.
    - \+ Low cost of possession:

- o  With about 0.7% RF modules failing per year and relatively easy and inexpensive repair, the possession costs of SSAs are in principle very low and, in any case, substantially lower than for klystron transmitters, due to the high cost of spare klystrons.
- SSA and klystrons have comparable power efficiencies. However, this must be analysed case by case.
  - +  For the ESRF booster's pulsed RF system, for instance, a reduction of the power consumption by a factor of almost 3 was obtained, thanks to possible capacitive filtering of the DC supply voltage.

This list of arguments is far from being exhaustive. The question of the best selection of technology to provide high RF power for accelerator projects is a standing item on the agenda of RF workshops and conferences. Yet, more and more labs are considering SSAs as an adequate solution for their RF needs. For the moment, I would state that for multi-megawatt applications, klystrons remain advantageous. However, for accelerator applications of several hundreds of kilowatts, in particular for medium-sized projects with limited human resources, solid state amplifiers will become more and more attractive.

## Acknowledgments

This paper is in tribute to Ti Ruan, who passed away in March 2014. In the early 2000s Ti Ruan initiated the design and the implementation of high power SSAs combining hundreds of transistors for larger accelerators. He is the father of the big SSAs implemented at SOLEIL, ESRF, and many other places around the world.

Many thanks are also due to the SOLEIL RF team: P. Marchand, R. Lopez, and F. Ribeiro; to the ELTA team: mainly J.-P. Abadie and A. Cauhepe; and to my RF colleagues at ESRF, in particular: J.-M. Mercier and M. Langlois. Their contributions constitute the backbone of this lecture.